\documentclass[12pt]{article}

\usepackage{latexsym, amssymb, a4, epsfig, graphicx, mathbbol, oldgerm}

\catcode`\@=11
\@addtoreset{equation}{section}

\catcode`@=12

\newcommand{\be}{\begin{equation}}
\newcommand{\ee}{\end{equation}}
\newcommand{\bea}{\begin{eqnarray}}
\newcommand{\eea}{\end{eqnarray}}
\newcommand{\bml}{\begin{mathletters} \baselineskip 10pt}
\newcommand{\eml}{\baselineskip 12pt \end{mathletters}}
\newcommand{\nn}{\nonumber}

\newcommand{\bra}{\langle}
\newcommand{\ket}{\rangle}

\newcommand{\Aphys}{\mathfrak{A}_{\mathrm{phys}}}

\newcommand{\pad}[2]{\frac{\partial #1}{\partial #2}}
\newcommand{\fud}[2]{\frac{\delta #1}{\delta #2}}

\newcommand{\vc}[1]{\mbox{\bf #1}}

\newcommand{\vcg}[1]{\mbox{\boldmath$#1$}}
\newcommand{\svcg}[1]{\mbox{\footnotesize\boldmath$#1$}}
\newcommand{\pb}[2]{\left\{#1 \, , \, #2 \right\}}
\newcommand{\spinor}[2]{\left[ \begin{array}{c}
                               #1 \\ #2 \end{array} \right]}

\newcommand{\tr}{\mbox{tr}}
\newcommand{\FP}{\mbox{FP}}
\newcommand{\ad}{\; \mbox{ad}}
\newcommand{\Ad}{\; \mbox{Ad}}
\newcommand{\e}{\mbox{e}}

\begin{document}

%title with preprint number
\title{\begin{flushright}
\normalsize FSU TPI 01/00
\end{flushright}
\vspace{2cm}
\bf \huge Instantons and Gribov Copies in the Maximally Abelian Gauge}
\vspace{5cm}

%title without preprint number
%\title{\bf \huge Instantons and Gribov Copies in the Maximally Abelian Gauge}
%\vspace{5cm}

\author{F.~Bruckmann, T.~Heinzl\thanks{Supported by DFG}, A.~Wipf \\
        Friedrich-Schiller-Universit\"at Jena\\ 
        Theoretisch-Physikalisches Institut \\ 
        Max-Wien-Platz 1, D-07743 Jena\\ 
        \and T.~Tok$^*$\\ 
        Universit\"at T\"ubingen\\ 
        Institut f\"ur Theoretische Physik\\ 
        Auf der Morgenstelle 14, D-72076 T\"ubingen}

\date{}

\maketitle

\thispagestyle{empty}

\begin{abstract}
We calculate the Faddeev-Popov operator corresponding to the maximally
Abelian gauge for gauge group $SU(N)$. Specializing to $SU(2)$ we look
for explicit zero modes of this operator. Within an illuminating toy
model (Yang-Mills mechanics) the problem can be completely solved and
understood. In the field theory case we are able to find an analytic
expression for a normalizable zero mode in the background of a single
`t~Hooft instanton. Accordingly, such an instanton corresponds
to a horizon configuration in the maximally Abelian gauge. Possible
physical implications are discussed.  
\end{abstract}

\newpage

%\include{toc}
%\include{habil1}
%\include{habil2}
%\addcontentsline{toc}{chapter}{Bibliography}

%\bibliographystyle{h-physrev}
%\bibliography{heinzl}

\section{Introduction}

The configuration space $\mathfrak{A}$ of gauge theories is a
``bigger-than-real-life-space'' \cite{kuchar:86a}. This is due to the
fact that the action of the gauge group $\mathfrak{G}$ relates
physically equivalent configurations along the gauge
orbits. Therefore, this action has to be divided out. In principle,
this division leads to the \emph{physical} configuration space,
$\Aphys = \mathfrak{A} / \mathfrak{G}$. In practice, however, this
division is not easily performed. The most efficient method to do so
is gauge fixing, where a subset of $\mathfrak{A}$ is identified with
$\Aphys$. This subset is characterized by choosing some condition on
the gauge potentials $A$ of the form $\chi [A] = 0$. Prominent
examples are the covariant gauge, $\chi_{\mathrm{cov}} = \partial_\mu
A_\mu$, or the axial gauge, $\chi_{\mathrm{ax}} = n \cdot A$. One
hopes that this condition satisfies both the requirements of existence
and uniqueness. Existence means that the hypersurface $\Gamma: \chi =
0$ intersects every orbit, while uniqueness requires that it does so
once and only once. It has first been shown by Gribov that the latter
requirement cannot be satisfied for non-Abelian gauge theories in the
covariant and Coulomb gauge \cite{gribov:78}. Shortly afterwards,
Gribov's observation has been proven for a large class of continuous
gauge fixings \cite{singer:78}. In the physics community, the lack of
uniqueness has become known as the Gribov problem. This just
paraphrases the difficulty in constructing the physical configuration
space which, by definition, is void of any (residual) gauge (or
Gribov) copies.

In order to analyse this issue it has turned out useful to describe
the gauge fixing not simply by a condition $\chi = 0$. Instead, in
order to study the global aspects of the problem, one formulates the
gauge fixing procedure in terms of a variational principle
\cite{yaffe:79,wilson:80,thooft:81a,semenov-tyan-shanskii:82,weisberger:83}.
To this end one tries to define an `action' functional $F$ in such a
way that the associated `classical trajectories' are nowhere parallel
to the orbits so that their union defines a gauge fixing
hypersurface. By this construction one completely suppresses
fluctuations in gauge directions which in the unfixed formulation do
not cost energy (or action) and thus make the path integral
ill--defined. Of course, by conservation of difficulties, one cannot
avoid the Gribov problem this way. 

The variational approach to gauge fixing has mainly been studied for
background type gauges like the Coulomb gauge, where one can indeed
construct a functional $F[A; U]$ with the following generic
properties: the critical points of $F$ along the orbits generated by
$U$ are the potentials $A$ satisfying the Coulomb gauge condition,
$\partial_i A_i = 0$. The Hessian of $F$ at these points is the
Faddeev-Popov operator FP. The Gribov region $\Omega^0$ is defined as
the set of transverse gauge fields for which det~FP is positive. This
is the set of relative minima of $F$. Its boundary $\partial \Omega^0$
is the Gribov horizon, where, accordingly, det FP = 0 because the
lowest eigenvalue of FP changes sign. It has been shown that $\Omega^0
\subset \mathfrak{A}$ is convex
\cite{semenov-tyan-shanskii:82,zwanziger:82a,zwanziger:82b}. This is
basically due to the linearity of FP in $A$
\cite{vanbaal:92}. Contrary to early expectations the Gribov region
still contains Gribov copies
\cite{semenov-tyan-shanskii:82,dell'antonio:90,dell'antonio:91}. Only
if one restricts to the set $\Lambda$ of \emph{absolute} minima and
performs certain boundary identifications, one ends up with the 
physical configuration space (also called the fundamental modular
domain) \cite{vanbaal:92}. 

As stated above, the appearance of horizon configurations $A \in
\partial \Omega^0$ implies that the gauge is not uniquely fixed; in
other words, there are gauge fixing degeneracies. Somewhat symbolically,
this can be shown as follows. Let  $A \in \Gamma$ have the infinitesimal
gauge variation $\delta A = D[A] \, \delta \phi$, $D$ denoting the
covariant derivative.  To check whether the gauge transform $A +
\delta A$ also satisfies the gauge condition, one calculates
\be
  \chi[A + \delta A] = \chi [A] + \fud{\chi}{A} \fud{A}{\phi} \, \delta
  \phi \equiv N [A] \cdot D[A] \, \delta \phi \; .
\ee
Here we have used that $\chi[A] = 0$ and defined the normal $N$ to the
gauge fixing hypersurface $\Gamma$. Now, if $A + \delta A$ is also in
$\Gamma$ we see that the FP operator,
\be
\label{FPND}
  \FP [A] \equiv N[A] \cdot D[A] \; , 
\ee
must have a zero mode given by the infinitesimal gauge transformation
$\delta \phi$. In this case, there are two gauge equivalent fields $A$
and $A + \delta A$ on $\Gamma$ and $A$ is a horizon
configuration. From (\ref{FPND}) one infers that there are two generic
reasons for this to happen. First, $\delta \phi$ can be a zero mode
already of $D[A]$. As the latter can be viewed as the `velocity' of a
fictitious motion along the orbits, its vanishing (on $\delta \phi$)
corresponds to a fixed point under the action of the gauge group. In
this case, the configuration $A$ is called reducible
\cite{singer:78,babelon:81b}. Obviously, these are always horizon
configurations. One might speculate whether there is a gauge fixing
such that reducible configurations are the \emph{only} horizon
configurations \cite{heinzl:99}. The second possibility for det FP to
vanish is that `orbit velocity' $D$ and normal $N$ are orthogonal,
which means that a particular orbit is tangent to $\Gamma$. This is
what usually happens for background type gauges like the Coulomb gauge
where $N$ is constant, i.e. independent of $A$.

In general, it is very hard to explicitly find horizon
configurations. For this reason, one has to concentrate on rather simple
and/or symmetric gauge potentials.  Again, the case best studied is
the Coulomb gauge. It is known that there are Gribov copies of
the classical vacuum $A=0$
\cite{vanbaal:92,jackiw:78,bender:78}. An even simpler example of
Gribov copies is provided by constant Abelian gauge fields on the
torus (the torons) \cite{vanbaal:92,koller:88}.  Configurations with a
radial symmetry have been discussed in the original work of Gribov
\cite{gribov:78}. An explicit example with axial symmetry has been
given by Henyey \cite{vanbaal:92,henyey:79}. 

On the lattice, the detection of Gribov copies has been reported for
the first time in \cite{forcrand:91}. It turns out that some of these
copies are lattice artifacts while others survive in the continuum
limit \cite{forcrand:95}. In a sense, therefore, the Gribov problem
becomes even more pronounced upon gauge fixing on the lattice. This is
of particular relevance for the lattice studies of the dual
superconductor hypothesis of confinement
\cite{parisi:75,mandelstam:76,thooft:76a}, where one mainly uses (a
lattice version \cite{kronfeld:87a}) of `t~Hooft's maximally Abelian
gauge (MAG) \cite{thooft:81a}. In order to extract physical results
within this approach one clearly has to control the influence of
Gribov copies. Finding the critical points of the lattice gauge fixing
functional is similar to a spin glass problem due to the high degree
of degeneracy. The difficulties in numerically determining the
absolute maximum\footnote{The maximization of the lattice functional
corresponds to a minimization of the continuum functional.} of the
lattice functional lead to an inaccuracy in observables of the order
of 10 \% \cite{bali:96,hart:97}.

The Gribov problem for the MAG so far has not been discussed in the
continuum. The purpose of this paper is to (at least partly) fill this
gap. The MAG and its defining functional will be reviewed in
Section~2. The Hessian of this functional is the FP operator which is
calculated for gauge group $SU(N)$. In Section~3 we specialize to
$SU(2)$ and give general arguments showing the existence of a Gribov
horizon. To provide some intuition, Section~4 introduces a simple toy
model for which the FP operator and determinant can be calculated
exactly. The presence of Gribov copies is shown explicitly. Finally,
in Section~5, we return to field theory and calculate the FP operator
in the background of a single instanton (in the singular
gauge). Again, we find an analytic expression for a normalizable zero
mode which shows that the single instanton is a horizon configuration
in the MAG. Some technical issues are discussed in Appendices A to D.

\section{The Maximally Abelian Gauge}

As explained in Appendix A, we decompose the gauge potential $A$ into
diagonal ($A^\parallel \in \mathcal{H}^\parallel$) and off-diagonal
($A^\perp \in \mathcal{H}^\perp$) components, $A = A^\parallel +
A^\perp$. The MAG is then defined by minimizing the following
functional
\be
  \label{GFF1}
  F[A ; U] \equiv \| (^{U \!\!}A)^\perp \|^2 \; .
\ee
$F$ is thus a functional of both the gauge field $A$ and the gauge
transformation $U \in SU(N)$. Via the parametrization
\be
  U (\phi) = \exp(- i \phi) = \exp (- i \phi^a T^a) \; , \quad \phi
\in su(N) \; ,
\ee
$F$ equivalently can be viewed as depending on the argument $\phi$ of
$U$.  The action of $U$ on $A$ is 
\be
  ^{U\!\!}A_\mu = U^{-1} A_\mu U + i U^{-1} \partial_\mu U \; .
\ee
With $F$ of (\ref{GFF1}) we are thus minimizing the `charged'
component $A^\perp$ along its orbit, which, roughly speaking, amounts
to maximizing the Abelian or `neutral' component $A^\parallel$. Hence
the name `maximally Abelian gauge'.

The Yang-Mills norm in (\ref{GFF1}) is the same as in the Yang-Mills
action and induced by the scalar product (\ref{YM_SCAL_PROD}),
\be
\label{NORM}
 \| A \|^2 \equiv \bra A, A \ket \equiv \int d^d x \, \tr A^2 \; .
\ee
Note that our conventions are such that this norm is positive for
\emph{hermitian} gauge fields $A$ with values in $su(N)$. The norm
(\ref{NORM}) can be viewed as the distance (squared) between $A$ and
the zero configuration $A = 0$. As the space $\mathfrak{A}$ of gauge
potentials is affine, the norm is gauge invariant in the following
sense,
\be 
  \| A - B \| = \| ^{U\!\!}A - \,  ^{U\!\!}B \| 
\ee
If the configuration $B$ is kept fixed, however, the norm ceases to be
gauge invariant and explicitly depends on $U$ or $\phi$. The same is
thus true for $F$ which accordingly changes along the orbit of
$A$ unless there is some (residual) invariance. For the functional
(\ref{GFF1}) such an invariance can indeed be found. Let $V = \exp (- i
\theta^\parallel)$ be an Abelian gauge transformation and consider
\be
  F[A; V] = \| (^{V\!\!}A)^\perp \|^2 = \| (V^{-1} A^\parallel V +
  V^{-1} A^\perp V + i V^{-1} dV)^\perp \|^2  \; .
\label{FAV1}
\ee
As $V$ is Abelian, the first and last terms on the r.h.s.~of
(\ref{FAV1}) vanish due to the projection on $\mathcal{H}^\perp$, and
we are left with
\be
  F [A; V] = \| (V^{-1} A^\perp V)^\perp \|^2 \; .
\label{FAV2}
\ee
At this point it is crucial to note that $V^{-1} A^\perp V$ is in
$\mathcal{H}^\perp$,  
\be
  \tr (H_i V^{-1} A^\perp V ) = \tr (V H_i V^{-1} A^\perp) = \tr (H_i
  A^\perp) = 0   \; .
\ee
Therefore, we can write for (\ref{FAV1}),
\be
  F [A; V] = \| V^{-1} A^\perp V \|^2 = \| A^\perp \|^2 = F [A; \Eins]
\; .
\ee
This immediately leads to the following Abelian invariance of $F$, 
\be
  F [A; VU] = F [^{U\!\!}A ; V] = F [^{U\!\!}A ; \Eins] = F [A; U] \; .
\label{ABEL_INV}
\ee
Note that our notation is such that $U$ acts prior to $V$, i.e.
\be
  ^{VU\!\!}A = (UV)^{-1} A \, UV + i (UV)^{-1} d(UV) \; .
\ee
Roughly speaking, the Abelian invariance implies that $F$ can be
thought of as some kind of `mexican hat' with the residual symmetry
corresponding to (Abelian gauge) rotations around its symmetry
axis. Accordingly, the Hessian of $F$ will have trivial zero modes
asssociated with the constant directions of $F$.

We are interested in the behaviour of $F[A; U]$ around the point $U =
\Eins$, i.e.~$\phi=0$, on the orbit of $A$. To this end we 
Taylor expand $F$ as
\be
  F [A; U] \equiv F [A; \phi] = F [A; 0] + \bra F^\prime [A; 0], \phi
  \ket + \frac{1}{2} \bra \phi , F^{\prime \prime} [A; 0] \, \phi \ket
  + O(\phi^3)
\ee
In order to do so we need the gauge transform $^{U\!\!}A$ as a power
series in $\phi$. The former can easily be found from
(\ref{GAUGETR_EXPL1}) and (\ref{GAUGETR_EXPL2}) with the result
\bea
  \label{TAYLOR_GT} ^{U\!\!}A_\mu &=& A_\mu + \frac{\exp(i \ad \phi) -
  \Eins}{i \ad \phi} \, (D_\mu \phi) \nn \\ 
  &=& A_\mu + D_\mu \phi +
  \frac{i}{2} [\phi , D_\mu \phi] + \frac{i^2}{3!} \Big[ \phi , [\phi
  , D_\mu \phi] \Big] + \ldots \; .
\eea
Not surprisingly, the covariant derivative $D_\mu = \partial_\mu - i
\ad (A_\mu)$ with $\ad (A) \, B \equiv [A,B]$ appears at this stage.
Inserting (\ref{TAYLOR_GT}) into (\ref{GFF1}) we obtain
\be 
\label{FTAYLOR}
  F [A; \phi] = \| A_\mu^\perp \|^2 + 2 \,  \bra A_\mu^\perp
  , (D_\mu \phi)^\perp \ket + \bra (D_\mu \phi)^\perp , (D_\mu
  \phi)^\perp \ket + i \, \bra A_\mu^\perp, [\phi , D_\mu \phi]^\perp \ket
  + \ldots \; .
\ee
In the following we are going to evaluate this expression term by
term. This requires some preparations. We will need the commutator
identity,
\be
  \bra A , [B,C] \ket = \bra B, [C, A] \ket = \bra C, [A,B] \ket   \; , 
\ee
which follows straightforwardly from the definition of the scalar
product. The latter equation shows that both the operator $\ad (A)$
and the covariant derivative $D[A]$ are anti-hermitean,
\bea
  \bra \phi, \ad (A) \psi \ket &=& - \bra \ad (A) \phi , \psi \ket \;
  , \label{AD_ADJ} \\
  \bra \phi, D[A] \psi \ket &=& - \bra D [A] \phi , \psi \ket \; .
\eea
The last two identities allow for an evaluation of the first
derivative $F^\prime$,
\be
  \bra A_\mu^\perp , (D_\mu \phi)^\perp \ket = \bra A_\mu^\perp ,
  D_\mu \phi  \ket =   - \bra D_\mu^\parallel A_\mu^\perp , \phi \ket
  = - \bra D_\mu^\parallel A_\mu^\perp , \phi^\perp \ket \; ,
\ee
with $D_\mu^\parallel \equiv \partial_\mu - i \ad A_\mu^\parallel$. We
thus have, to first order in $\phi$,
\be 
  F [A; \phi] = \| A_\mu^\perp \|^2 - 2 \, \bra D_\mu^\parallel
  A_\mu^\perp , \phi^\perp \ket + O(\phi^2) \; .
\ee
Note that to this order, $F$ does not depend on the Cartan component
$\phi^\parallel$.  We immediately read off the critical points defining
the MAG,
\be
\label{MAGF}
  D_\mu^\parallel A_\mu^\perp \equiv  D_\mu A_\mu^\perp = 0 \; .
\ee
The second derivative requires considerably more efforts. We relegate
the explicit calculations to Appendix~\ref{app:C}, where we obtain for
the Taylor expansion of $F [A; \phi]$,
\bea
\label{F12}
  F [A; \phi] &=& \| A_\mu^\perp \|^2 - 2 \bra D_\mu^\parallel
  A_\mu^\perp , \phi^\perp \ket + i \bra \phi^\perp , \ad (D_\mu
  A_\mu^\perp ) \phi^\parallel \ket\nn \\ 
  &-& \bra \phi^\perp , \Big[
  D_\mu^\parallel D_\mu ^\parallel + \ad^2 A_\mu^\perp - i (\ad
  A_\mu^\perp ) \mathbb{Q} D_\mu - i \ad (D_\mu^\parallel A_\mu^\perp
  ) \Big] \phi^\perp \ket \nn \\ 
  &+& O(\phi^3) \; .
\eea
Here we have defined a projection $\mathbb{Q}$ onto the complement
$\mathcal{H}^\perp$ of the Cartan subalgebra such that $\mathbb{Q}
\phi = \phi^\perp$.  The term in (\ref{F12}) depending on
$\phi^\parallel$ may seem somewhat strange but is actually necessary
to guarantee the Abelian invariance (\ref{ABEL_INV}). It vanishes on
the gauge fixing hypersurface $\Gamma$ defined by (\ref{MAGF}).

From (\ref{F12}) we can easily read off the Faddeev-Popov operator
which is the Hessian of $F$ evaluated on $\Gamma$ (i.e.~at the
critical points),
\be
\label{FPMAG_GEN}
  \FP = - \mathbb{Q} \left( D_\mu^\parallel D_\mu ^\parallel + \ad^2
  A_\mu^\perp - i (\ad A_\mu^\perp ) \mathbb{Q} D_\mu \right) \mathbb{Q}
  \; .
\ee
In effect we have performed a saddle point approximation to the
functional $F [A; \phi]$. The `equation of motion' is the gauge fixing
condition, and the fluctuation operator is the FP operator. In this
approximation the functional on $\Gamma$ reads
\be
  F [A; \phi] = \| A_\mu^\perp \|^2 + \bra \phi , \FP \phi \ket +
  O(\phi^3) \; .
\ee
As stated in the introduction, it is in general rather difficult (in a
continuum formulation) to find explicit examples of Gribov copies. The
MAG is no exception from this rule. The nontrivial task is to find
normalizable zero modes of FP given by (\ref{FPMAG_GEN}) which is a
complicated partial differential operator.  We are, however,
encouraged by lattice calculations, in which such copies have been
detected numerically, for the first time in \cite{hioki:91a} and with
refined techniques in \cite{bali:96,hart:97}.
One should keep in mind, though, that some (if not all) of these
copies can be lattice artifacts which do not survive in the continuum
limit. To study the possible appearance of Gribov copies in the
continuum we have to perform several simplifications. The first one
will be to consider the case of gauge group $SU(2)$.

\section{The FP Operator for $\vcg{SU(2)}$ --- General Considerations}

For $SU(2)$, the gauge fixing condition (\ref{MAGF}) of the MAG can be
rewritten in terms of the gauge field components $A_\mu^3 \in
\mathcal{H}^\parallel$ and $A_\mu^{\pm} \in \mathcal{H}^\perp$, 
\be
\label{MAGFIX}
  (\partial_\mu \pm i A_\mu^{3}) A_\mu^\pm = 0 \; , \quad A_\mu^\pm
  \equiv A_\mu^1 \pm i A_\mu^2 \; . 
\ee
The fact that these are only \emph{two} requirements already implies
(by counting of degrees of freedom) that there remains a residual
gauge freedom corresponding to a one-dimensional subgroup which can
only be $U(1)$. Superficially, the gauge fixing looks like a
background gauge which would actually be true if the neutral component
$A_\mu^3$ were independent of the charged one, $A_\mu^\perp$. As
these, however, are two components of one and the same configuration
they are not independent, and the gauge fixing condition is quadratic,
i.e.~\emph{nonlinear} in $A_\mu$. This makes life somewhat complicated
(although it does not spoil the renormalizability of the gauge
\cite{min:85}). A BRST approach, for example, necessitates the
introduction of four-ghost terms. In a path integral formulation,
these ghost interactions `regularize' the usual bilinear FP ghost term
in the presence of zero modes \cite{schaden:99}.

The FP operator for $SU(2)$ simplifies considerably as the last term
in (\ref{FPMAG_GEN}) vanishes. One is thus left with the following sum
of two operators,
\be
\label{FPMAG_SU2}
  \FP = - \mathbb{Q} \Big(D_\mu^\parallel D_\mu^\parallel + \ad^2
  (A_\mu^\perp) \Big) \mathbb{Q} \; .
\ee
Using the notation (\ref{REIN_NOT}), FP can be viewed as a 3 $\times$ 3
matrix in color space. The operator $\mathbb{Q}$ projects onto the two
directions perpendicular to the $z$-axis so that the third row and
column of FP vanish identically. The associated trivial zero mode
corresponds to the residual $U(1)$ gauge freedom which remains unfixed
by the MAG. Explicitly, one has for the nonvanishing entries of FP, 
\bea
  (D_\mu^\parallel D_\mu^\parallel)^{\bar a \bar b} &=& \delta^{\bar a
  \bar b} (\Box - A_\mu^3 A_\mu^3) - \epsilon^{\bar a \bar b}
  (\partial_\mu A_\mu^3 + 2 A_\mu^3 \partial_\mu) \; ,\label{REIN_REP1} \\
  \big( \ad^2 (A_\mu^\perp)\big)^{\bar a \bar b} &=& 
  \delta^{\bar a \bar b} A_\mu^{\bar c}  A_\mu^{\bar c} - A_\mu^{\bar
  a} A_\mu^{\bar b} \; . \label{REIN_REP2}
\eea
Summing these two terms leads to the representation of FP given in
equation (12) of \cite{quandt:98a}\footnote{Note, however, that in
this reference the gauge potentials are defined as being
\emph{anti-hermitean.}}. 

Being (the negative of) a Laplacian, the operator $ - D_\mu^\parallel
D_\mu^\parallel$ is nonnegative. The same is true for $\ad (A_\mu^\perp)
\ad (A_\mu^\perp)$ as will be shown in what follows. We define the
\textit{hermitean} matrix $C$ via
\be
  [A_\mu^\perp , \phi^\perp] \equiv i C \; , 
\ee
and calculate, using (\ref{AD_ADJ}),
\bea
  \bra \phi, \mathbb{Q} \ad (A_\mu^\perp) \ad (A_\mu^\perp) \, \mathbb{Q}
  \, \phi \ket 
  &=& - \bra \ad{A_\mu^\perp} \phi^\perp , \ad{A_\mu^\perp} \phi^\perp
  \ket \\
  &=& - \bra iC , iC \ket = \bra C , C \ket \ge 0 \; . 
\eea
One can as well use the representations (\ref{REIN_REP1}),
(\ref{REIN_REP2}) and the Cauchy--Schwarz inequality to end up with
the same result. The $SU(2)$ FP operator from (\ref{FPMAG_SU2}) is
thus the \textit{difference of two positive semidefinite operators}
which we abbreviate for the time being as
\be
\label{FPAB}
  \FP = A - B \; , \quad A,B \ge 0 \; .
\ee
The inequality denotes the fact that $A$ and $B$ have nonnegative
spectrum. The identity (\ref{FPAB}) already suggests that if $B$ is
`sufficiently large', FP will develop a vanishing eigenvalue. Let us
make this statement slightly more rigorous. To this end we modify an
argument used in \cite{babelon:81b,creutz:79} for background type
gauges.

\begin{figure}
  \begin{center}
    \epsfig{figure=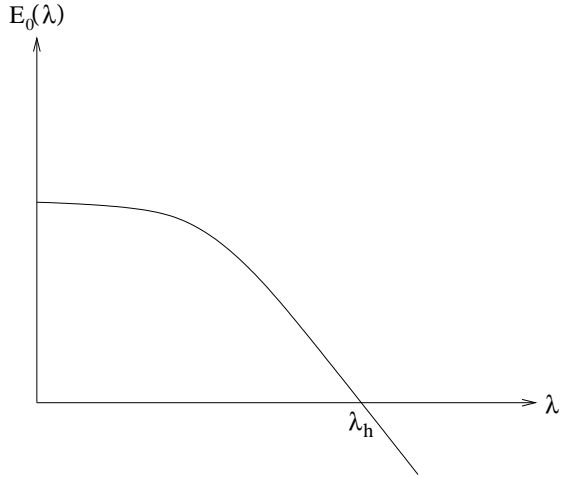,scale=0.7,angle=0}
  \end{center}
\caption{Qualitative behavior of the lowest eigenvalue of FP as a
function of the `flow parameter' $\lambda$. The parameter value
$\lambda_h$ corresponds to a horizon configuration.}
\label{fig:SFLOW}
\end{figure}

First of all we note that together with the configuration
$(A^\parallel , A^\perp)$ also the scaled configuration $(A^\parallel
, \lambda A^\perp)$, with $\lambda$ some (positive real) parameter,
will be in the MAG. The associated FP operator is
\be
  \FP [A^\parallel , \lambda A^\perp] \equiv \FP (\lambda) = A -
  \lambda^2  B \; . 
\ee
Let us denote the lowest eigenvalue and the associated eigenfunction
of $\FP(\lambda)$ by $E_0 (\lambda)$ and $\phi_0 (\lambda)$,
respectively, 
\be
  \FP (\lambda) \,  \phi_0 (\lambda)  = E_0 (\lambda) \,  \phi_0
  (\lambda)  \; .
\ee
From (\ref{FPAB}) one must have $E_0 (0) \ge 0$. If we turn on
$\lambda$, a straightforward application of the Hellmann--Feynman
theorem leads to
\be
  \pad{}{\lambda} E_0 (\lambda) = - 2 \lambda \,  \bra \phi_0
  (\lambda) , B  \, \phi_0 (\lambda) \ket  \le 0 \; , 
\ee
whence the function $E_0 (\lambda)$ has negative slope. In addition, it
has to be concave \cite{galindo:91}\footnote{It is exactly for this
reason that the second order perturbation theory correction to any
groundstate is always negative.} so that, for $\lambda$ sufficiently
large, there will be a zero-mode at some value, say $\lambda_h$ (see
Fig.~\ref{fig:SFLOW}). In a way we have thus determined a `path'
within the MAG fixing hypersurface that leads us from the interior of
the Gribov region ($\lambda = 0$) to its boundary ($\lambda =
\lambda_h$).

As a result we can state that generically there have to be Gribov
copies within the MAG if the non-diagonal components $A^\perp$ of the
gauge fields become sufficiently large.

\section{A Toy Model}

In order to have an illustration of the somewhat abstract notions of
the preceding sections we will analyse an example with a finite number
of degrees of freedom~\cite{pause:98}. To this end we employ a
Hamiltonian formulation in $d$ = 2 + 1 and consider only gauge
potentials $A_\mu$ which are spatially constant. Renaming $A_i^a =
x_i^a$, $i = 1,2$, $a = 1,2,3$, the Lagrangian becomes
\be
\label{TOYLAG}
  L = \frac{1}{2} (D_0^{ab} x_i^b)^2 \equiv \frac{1}{2} (\dot x_i^a -
  \epsilon^{abc}A_0^c x_i^b)^2 \; .
\ee
One way of arriving at this Lagrangian is by gauging a free particle
Lagrangian $L_0 = \dot x_i^a \dot x_i^a / 2$ via minimal substitution,
i.e.~by replacing the ordinary time derivative $\partial_0$ with the
covariant derivative $D_0^{ab}$. To keep things as simple as possible,
we have not introduced any (Yang-Mills type) interaction; we are
anyhow only interested in the kinematics of the problem.

\begin{figure}
  \begin{center}
    \epsfig{figure=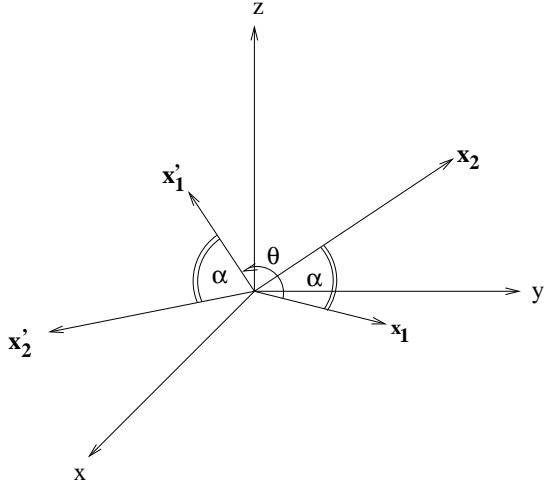,scale=0.7,angle=0}
  \end{center}
\caption{An isospace (gauge) rotation (by an angle $\theta$) in the toy
model, transforming the configuration $(x_1 , x_2) \to (x_1^\prime ,
x_2^\prime)$. The lengths of the vectors and the angle $\alpha$
inbetween them are invariant.}
\label{fig:VECTORS}
\end{figure}

Defining the canonical momenta $p_i^a = D_0^{ab} x_i^b$, the
Lagrangian (\ref{TOYLAG}) can be recast in first order form
\be 
  L = p_i^a \dot x_i^a - \frac{1}{2} p_i^a p_i^a + A_0^a G^a \; , 
\ee
where we have introduced the operator $G^a$ leading to Gauss's
law
\be
  G^a \equiv \epsilon^{abc} x_i^b p_i^c \equiv D_i^{ab} p_i^b = 0 \; .
\ee
Obviously, $G^a$ is the total angular momentum of two point particles
in $\mathbb{R}^3$ (= color isospace) with position vectors $\vcg{x}_1$
and $\vcg{x}_2$. Gauge transformations are thus $SO(3)$ rotations of
these vectors which do not change their relative orientation (i.e. the
angle $\alpha$ inbetween them). This is illustrated in
Fig.~\ref{fig:VECTORS}.

As usual we will work in the Weyl gauge, $A_0 = 0$, so that Gauss's
law has to be imposed `by hand', and, after quantization, holds upon
acting on physical states. Once the Weyl gauge has been chosen, there
still is the freedom of performing time independent gauge
transformations. This will be (partially) fixed using the MAG. For the
case at hand, there are several equivalent ways of formulating the
latter.

To avoid writing too many indices we denote $\vcg{x}_1 \equiv
\vcg{x} = (x,y,z)$, $\vcg{x}_2 \equiv \vcg{X} = (X,Y,Z)$. An arbitrary
vector $\vcg{A}$ will be decomposed according to
\bea
\label{PP_DECOMP}
  \vcg{A}_\parallel &\equiv& A_z \vcg{e}_z \; , \\
  \vcg{A}_\perp &\equiv&  A_x \vcg{e}_x + A_y \vcg{e}_y \; , 
\eea
which represents the decomposition into Cartan (= $z$) component and its
complement. The MAG condition then reads explicitly
\be
\label{TOYMAG1}  
  \chi^a \equiv D_i^{ab} (x_{i \parallel}) x_{i\perp} =
  \epsilon^{abc} x_{i\parallel}^b x_{i \perp}^c = 0 \; ,
\ee
or, in components,
\bea
  \chi^1 &=& -yz - YZ = 0 \; , \nn \\
  \chi^2 &=&  xz + XZ = 0 \; , \label{TOYMAG2} \\
  \chi^3 &=& 0 \nn \; . 
\eea
The last condition is just an empty tautology so that there are in
fact only \textit{two} gauge conditions\footnote{In Dirac's
terminology \cite{dirac:64}, $\chi^3$ is \textit{strongly} zero and
thus does not contribute in the calculation of any Poisson
bracket.}. Of course, this just corresponds to the fact that the gauge
rotations generated by $G^3$ (the rotations around the $z$--axis)
remain unfixed (cf.~the remark after (\ref{MAGFIX})).

\begin{figure}
  \begin{center}
    \epsfig{figure=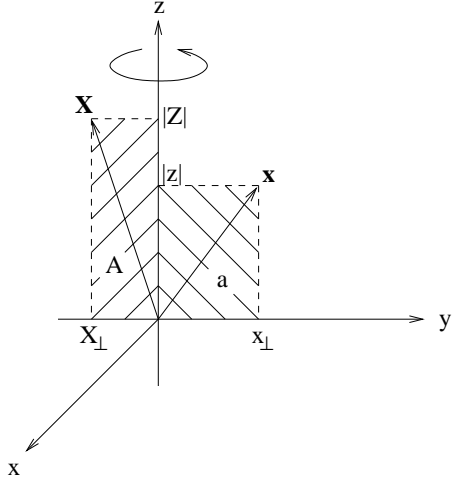,scale=0.7,angle=0}
  \end{center}
\caption{The MAG condition in the toy model. The areas $A$ and $a$
have to be the same. We have arbitrarily chosen $\vcg{x}$ and
$\vcg{X}$ to lie in the $yz$--plane. The residual $U(1)$ gauge freedom
corrresponds to rotations around the $z$--axis.}
\label{fig:AREAS}
\end{figure}

The MAG conditions (\ref{TOYMAG2}) can be easily visualized. The
projections $\vcg{x}_\perp$ and $\vcg{X}_\perp$ have to be collinear
with their magnitudes being related through
\be
\label{TOYMAG3}
  |z| \,  x_\perp = |Z| \, X_\perp \; .
\ee
The MAG is thus obtained by rotating the configuration (${\vcg{x},
\vcg{X}}$) in such a way that both vectors are as close to the $z$-axis
as possible. This is achieved as shown in Fig.~\ref{fig:AREAS}. 
$\vcg{x}$ and $\vcg{X}$ are the diagonals of two rectangles with sides
$|z|$, $x_\perp$ and $|Z|$, $X_\perp$, respectively. If the areas $a$
and $A$ of the rectangles coincide, $a = A$, the configuration is in
the MAG. Algebraically, the notion of being `close to the $z$--axis'
is measured by the function
\be
  F(\vcg{x}, \vcg{X}) \equiv x_\perp^2 + X_\perp^2 \; .
\ee
One can easily show that the conditions (\ref{TOYMAG1}) or
(\ref{TOYMAG2}) minimize $F$ and thus make the `nondiagonal'
components of $\vcg{x}$ and $\vcg{X}$ as small as possible. We mention
in passing that the trivial solution of (\ref{TOYMAG2}) given by $z =
Z = 0$ corrresponds to a maximum of $F$ so that we can always assume
$z$ or $Z \ne 0$ (except for the zero--configuration representing the
origin).

It is obvious from Fig.~\ref{fig:AREAS} that rotations around
the $z$--axis leave both $F$ and the MAG condition invariant and thus
correspond to a residual $U(1)$ gauge freedom.  As expected,
this situation is reflected in the FP operator,
\be
  \FP =  \left. \pb{\chi^a}{G^b} \right|_{\svcg{\chi} = 0} \; ,
\ee
which, in matrix notation, can be written as 
\be
  \FP = \left( \begin{array}{ccc}
                 z^2 + Z^2 - y^2 - Y^2 & xy + XY & 0 \\
                 xy + XY & z^2 + Z^2 - x^2 - X^2 & 0 \\
                 0 & 0 & 0 
               \end{array} \right) . 
\ee
The zero entries in the third row and column are a trivial consequence
of the residual $U(1)$ and correspond to the action of
the $\mathbb{Q}$--projection in (\ref{FPMAG_GEN}). The eigenvalues of
FP are found to be
\bea
  E_3 &=& 0 \; , \\
  E_+ &=& z^2 + Z^2 \; , \\
  E_- &=& z^2 + Z^2 - x_\perp^2 - X_\perp^2 \; .
\eea
Let us concentrate on the eigenvalues $E_\pm$ which are not related to
the residual Abelian gauge freedom. Configurations where one of these
vanishes are located on the Gribov horizon and reflect some
non-trivial residual gauge freedom different from the $U(1)$ above. A
particular (in some sense trivial) class of horizon configurations
consists in the reducible configurations as discussed in the
introduction. These have a higher symmetry than generic configurations
(a nontrivial stabilizer or isotropy group). In other words, they are
fixed points under the action of (a subgroup of) the gauge
group. Technically, they show up by inducing zero modes of the
Laplacian $\Delta^{ab} = D_i^{ac}D_i^{cb}$ (see Appendix~D). Within
our example, the reducible configurations are readily identified
\cite{pause:98,arms:81} by simple symmetry considerations.  The origin
is invariant under the whole action of $SO(3)$ while configurations
with $\vcg{x}$ and $\vcg{X}$ collinear are invariant under rotations
around their common direction which clearly corresponds to a
$U(1)$. This is nicely reflected in the spectrum of FP. At the origin,
both $E_\pm$ vanish, while a collinear configuration can always be
rotated in the $z$--axis so that its stabilizer coincides with the
standard residual $U(1)$ corresponding to $E_3 = 0$. This U(1)
stabilizer is thus `hidden' in the residual $U(1)$. Fixing the latter
by demanding e.g. $x = X = 0$, does, however, not affect
configurations collinear along the $z$--axis so that these will induce
zero modes of FP even after residual gauge fixing \cite{pause:98}.

\begin{figure}
  \begin{center}
    \epsfig{figure=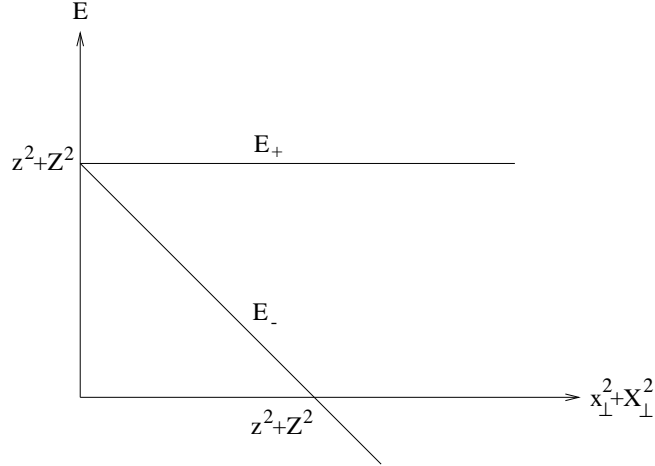,scale=0.7,angle=0}
  \end{center}
\caption{Behavior of the eigenvalues of FP in the toy model as a
function of the magnitude $x_\perp^2 + X_\perp^2$ of the
`nondiagonal' components. A zero mode arises when $x_\perp^2 +
X_\perp^2 = z^2 + Z^2$.}
\label{fig:EIGENVALS}
\end{figure}

There is a remaining possibility for a vanishing eigenvalue. While
$E_+$ is always positive, $E_-$ vanishes if $z^2 + Z^2 = x_\perp^2 +
X_\perp^2$. This happens for configurations where $\vcg{x}$ and
$\vcg{X}$ are of the same length and orthogonal to each
other. Elementary trigonometry implies that in this case the two areas
$a$ and $A$ are always the same, irrespective of the location of the
configuration relative to the $z$--axis. Thus, there is an additional
residual $U(1)$ gauge freedom for such exceptional
configurations. This can be nicely illustrated in terms of a `spectral
flow' as a function of $x_\perp^2 + X_\perp^2$ (see
Fig.~\ref{fig:EIGENVALS}). We thus have found an explicit realization
of the general results of Section 3, in particular of
Fig.~\ref{fig:SFLOW}.

\section{The FP Operator in an Instanton Background}

The natural question arising at this point is the following: is there a
way of extending the results of the toy model to the realistic field
theory case? The answer given in this section will be affirmative. 

Our motivation stems from the observation made by Brower et
al.~\cite{brower:97b} that the single `t~Hooft instanton \textit{both}
in the singular \textit{and} regular gauge satisfies the MAG condition
(\ref{MAGF}). For the instanton in the singular gauge\footnote{We use
the conventions of \cite{schaefer:98}.} (or `singular instanton', for
short) given by
\be
\label{INST_SING}
  A_\mu^{\mathrm{sing}} (x) = 2 \, \bar \eta_{\mu \nu}^a \,
  \frac{\rho^2}{x^2} \,  \frac{x_\nu}{x^2 + \rho^2} \, \sigma^a /2 \; ,
\ee
with $\rho$ denoting the instanton size, the MAG fixing functional $F$
is finite, while for the instanton in the regular gauge,
\be
\label{INST_REG}
  A_\mu^{\mathrm{reg}} (x) = 2 \, \eta_{\mu \nu}^a  \, \frac{x_\nu}{x^2 +
  \rho^2} \, \sigma^a / 2 \; , 
\ee
it diverges. The two configurations $A_\mu^{\mathrm{sing}}$ and
$A_\mu^{\mathrm{reg}}$ are related through the gauge transformation 
\be
\label{SINGREG}
  g(x) = \hat{x}_4 + i \, \hat{x}^a \sigma^a \; , 
\ee
where $\hat{x}_\mu = x_\mu / r$, $r = (x^2)^{1/2}$ denoting the
modulus of the Euclidean position $x$. If we adopt the point of view
that we have to take the minima of $F$ to define the Gribov region
$\Omega^0$ of the MAG then $A_\mu^{\mathrm{sing}}$ is located in
$\Omega^0$ while $A_\mu^{\mathrm{reg}}$ is not.  This is corroborated
by the quoted work of Brower et al.~\cite{brower:97b} which, when
translated into our language, amounts to the following.  One
numerically constructs a `path' $\gamma (R) \in \Gamma$ connecting
$A_\mu^{\mathrm{sing}}$ with $A_\mu^{\mathrm{reg}}$. Along this
path\footnote{In \cite{brower:97b} the parameter $R$ is the radius of a
monopole loop associated with the configuration $A_\mu (R)$ located on
$\gamma$ somewhere inbetween $A_\mu^{\mathrm{sing}}$ = $A_\mu (R=0)$
and $A_\mu^{\mathrm{reg}} = A_\mu (R = \infty)$.} (beginning at the
singular instanton) the MAG functional $F$ is monotonically
rising. The configurations $A_\mu (R)$ along the path are determined
by applying a (singular) gauge transformation $\Omega$ which takes the
singular instanton to $A_\mu (R)$, i.e. $A_\mu (R) = \, ^{\Omega
\!\!}A_\mu^{\mathrm{sing}}$.  Hence $\gamma (R)$ is a path \textit{both}
within $\Gamma$ \textit{and} the single instanton orbit. Accordingly,
there must be an infinitesimal gauge transformation of the singular
instanton that does not leave $\Gamma$ and thus must be a zero mode of
FP $[A^\mathrm{sing}]$. In what follows we will try to explicitly
determine this zero mode.

The first step of this program consists in the calculation of the FP
operator in the background of a singular instanton. Plugging
(\ref{INST_SING}) into (\ref{REIN_REP1}) and (\ref{REIN_REP2}) one
obtains the result
\be
  \FP^{\bar a \bar b} = - \delta^{\bar a \bar b} \Box + 2 \, 
  \epsilon^{\bar a \bar b} \, a(r) (x_2 \partial_1 - x_1 \partial_2 +
  x_3 \partial_4 - x_4 \partial_3) \; .
\ee
We have discarded the vanishing third row and column (resulting
from the action of $\mathbb{Q}$) and introduced the (singular)
instanton profile function,
\be
\label{PROFILE}
  a (r) = 2 \, \frac{\rho^2 / r^2}{r^2 + \rho^2} = 2 \left(\frac{1}{r^2}
  - \frac{1}{r^2 + \rho^2} \right) \; . 
\ee
We are looking for normalizable zero modes $\phi$ of the FP operator,
\be
\label{FPZM}
  \FP \, \phi = 0 \; , \quad \bra \phi , \phi \ket < \infty \; , 
\ee
where $\phi(x)$ now is a two-component vector (field) living in the
complement of the Cartan subalgebra. Solving the equation (\ref{FPZM})
for the zero mode is basically an exercise in group theory as will
become clear in a moment. If we define the generators of
four-dimensional Euclidean rotations as
\be
\label{LMUNU}
  L_{\mu \nu} = -i (x_\mu \partial_\nu - x_\nu \partial_\mu) \; , \quad
  \mu , \nu = 1, \ldots , 4 \; ,
\ee
the FP operator can be written in 2 $\times$ 2 matrix notation as
\be
  \FP = \left( \begin{array}{cc}
                   - \Box & - 2 i \, a(r) (L_{12} - L_{34}) \\
                   2i \, a(r) (L_{12} - L_{34}) & - \Box 
               \end{array} \right) 
\ee
It is straightforward to check that the $L_{\mu \nu}$ indeed satisfy
the Lie algebra of $SO(4)$. In analogy with the Lorentz group one
introduces the angular momentum and `boost' generators
\bea
  L_i &\equiv& \frac{1}{2} \epsilon_{ijk} L_{jk} \label{LK1} \; \\
  K_i &\equiv& L_{i4} \label{LK2} \; ,
\eea
and their linear combinations,
\bea
  M_i &\equiv& \frac{1}{2} (L_i - K_i) = - \frac{i}{2} \, \bar
  \eta_{\mu \nu}^i \, x_\mu \partial_\nu \; , \\
  N_i &\equiv& \frac{1}{2} (L_i + K_i) = - \frac{i}{2} \, 
  \eta_{\mu \nu}^i \, x_\mu \partial_\nu\; .
\eea
These can be viewed as the self-dual and anti-self-dual parts of
$L_{\mu \nu}$, if `duality' is understood as the exchange of $\vc{L}$
and $\vc{K}$.  The operators $M_i$ and $N_i$ generate two independent
$SU(2)$ subgroups with Casi\-mirs $M^2$ and $N^2$ having eigenvalues
$M(M+1)$ and $N(N+1)$, respectively \cite{ramond:90}. It is important
to note that $M$ and $N$ will in general be half-integer,
\be 
  M, N \in \{ 0, 1/2, 1, \ldots \} \; .
\ee
This fact is well known from the algebraic treatment of the hydrogen
atom which has a hidden dynamical $O(4)$ symmetry (see
e.g.~\cite{merzbacher:98}).  In addition, as FP is a 2$\times$2
matrix, it can be expanded in terms of Pauli matrices, so that
altogether we find the rather compact result,
\be
  \FP = - \Box \Eins + 4 a(r) M_3 \, \sigma_2 \; . 
\ee
Plugging this into (\ref{FPZM}) results in a four-dimensional
Schr\"odinger equation with spin having a high degree of symmetry. A
complete set of commuting observables is given by the Casimirs $M^2$
and $N^2$, their projections $M_3$ and $N_3$ (with eigenvalues $m$ and
$n$) and the Pauli matrix $\sigma_2$ (eigenvalues $ s = \pm
1$). Replacing $\sigma_2$ by its eigenvalue and rewriting the
Laplacian in terms of the radial coordinate $r$ we are left with
\be
  \FP(s) \equiv - \partial_r^2 - \frac{3}{r} \partial_r +
  \frac{2}{r^2} (\vc{M}^2 + \vc{N}^2) + 4 a(r) M_3 \, s 
\ee
This is indeed a 4$d$ radially symmetric Hamiltonian. Upon closer
inspection, the Casimir term turns out to become even simpler. Using
the representations (\ref{LMUNU}), (\ref{LK1}) and (\ref{LK2}) one
finds that
\be
  \vc{N}^2 - \vc{M}^2 = \vc{L} \cdot \vc{K} = 0 \; , 
\ee
so that FP finally becomes
\be 
  \FP (s) = - \partial_r^2 - \frac{3}{r} \partial_r + \frac{4}{r^2}
  \vc{M}^2 + 4 a(r) M_3 \, s \; .
\ee
The eigenfunctions of FP will therefore depend on the quantum numbers
$M \in \{0, 1/2, 1, \ldots \}$, $m,n \in \{ -M , -M + 1,  \ldots , M \}$
and $s = \pm 1$. Chosing the coordinates
\be
  x = r \, (\cos \theta \cos \varphi_{12} , \cos \theta \sin \varphi_{12} ,
  \sin \theta \cos \varphi_{34} , \sin \theta \sin \varphi_{34}) \; , 
\ee
with $0 \le \theta \le \pi/2$, $0 \le \varphi_{12}, \varphi_{34} \le 2 \pi$,
the eigenfunctions can be written as follows,
\be
  \phi = f_{Mm} (r) \, h_{Mmn} (\theta) \, y_{mn} (\varphi_{12}) \, z_{mn}
  (\varphi_{34}) \, \chi_s \; .
\ee
The $\chi_s$ are the eigenspinors of $\sigma_2$,
\be
  \chi_\pm = \frac{1}{\sqrt{2}} \spinor{1}{\pm i} \; .
\ee
The Schr\"odinger equation factorizes accordingly. Introducing the
dimensionless variable $R = r/\rho$ and defining a function $g(R)$ via
\be
  f(R) \equiv g(R)/R^{3/2} \; , 
\ee
(we omit the subscripts of $f$) the radial equation for the zero mode
becomes
\be
\label{GEQ1}
  \left[ - \partial_R^2 + \frac{4M(M+1) + 3/4}{R^2} + 
  \frac{8ms}{R^2 (1 + R^2)} \right] g(R) = 0 \; .
\ee
We are looking for a normalizable zero mode, or, in other words, a
bound state with vanishing energy. For this we need an attractive
potential. We thus must have $ms < 0$, and we choose $s=-1$, $m>0$ in
what follows. The bound state equation (\ref{GEQ1}) thus becomes
\be
\label{GEQ2}
  \left[ - \partial_R^2 + \frac{4M(M+1) - 8m + 3/4}{R^2} + 
  \frac{8m}{1 + R^2} \right] g(R) = 0 \; .
\ee
\begin{figure}[t]
  \begin{center}
    \epsfig{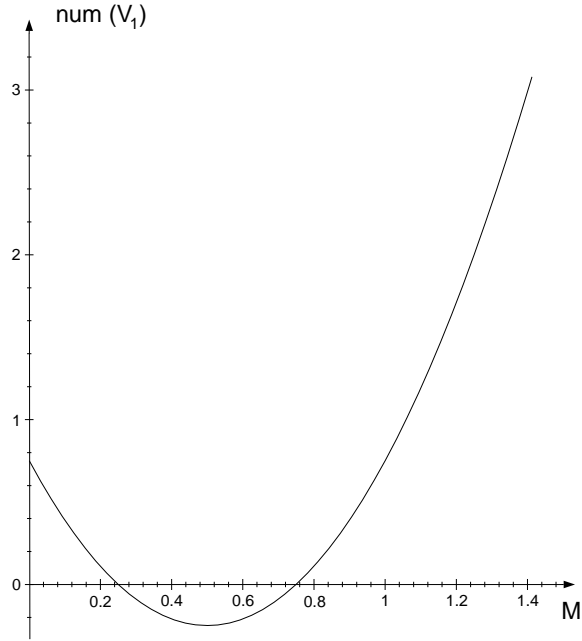}
  \end{center}
\caption{The numerator of the potential term $V_1$ as a function of
the quantum number $M$. The only (half-)integer leading to attraction
(negative $V_1$) is $M$ = 1/2.}
\label{fig:V1}
\end{figure}
This equation has already been obtained by Brower et
al.~\cite{brower:97b} in the stability analysis of their monopole
solutions. These authors, however, have overlooked the fact that $M$
is half-integer which is crucial for obtaining the correct
solution (see below). In addition they approximated the profile
function $a(r)$ by $1/r^2$ (in the limit of small monopole loops). We
will instead solve (\ref{GEQ2}) exactly.  The latter is an effective
one--dimensional Schr\"odinger equation with a Hamiltonian
\be
  H_R \equiv - \partial_R^2 + V_1 (R) + V_2 (R) \; .
\ee
The second potential term, $V_2$, is always positive (for $m>0$). Only
the first term, $V_1$ has a chance of becoming negative leading to
attraction. Technically, this is due to the relative minus sign in the
profile function (\ref{PROFILE}) of the singular instanton. In the
regular gauge, this is absent so that both $V_1$ and $V_2$ are
positive and there are no normalizable zero modes.  As $m$ is bounded
by $M$, the Casimir term $M(M+1)$ in (\ref{GEQ2}) will always win for
large $M$. We thus should make $M$ as small and $m$ as large as
possible. We thus take $m=M$ and plot the numerator of $V_1$ as a
function of $M$ (see Fig.~\ref{fig:V1}).  Obviously, there is exactly
one solution for $M$ which makes $V_1$ negative, namely $M = 1/2 =
m$. We have explicitly checked that for $M > 1/2$ there is no bound
state solution\footnote{The claim of Brower et al.~\cite{brower:97b},
that attraction occurs for $m = 1$ with the ground state having $M=1$,
thus cannot be substantiated.}.  The associated potential $V_1 + V_2$
is plotted in Fig.~\ref{fig:POTENTIAL}.
\begin{figure}[t]
  \begin{center}
    \epsfig{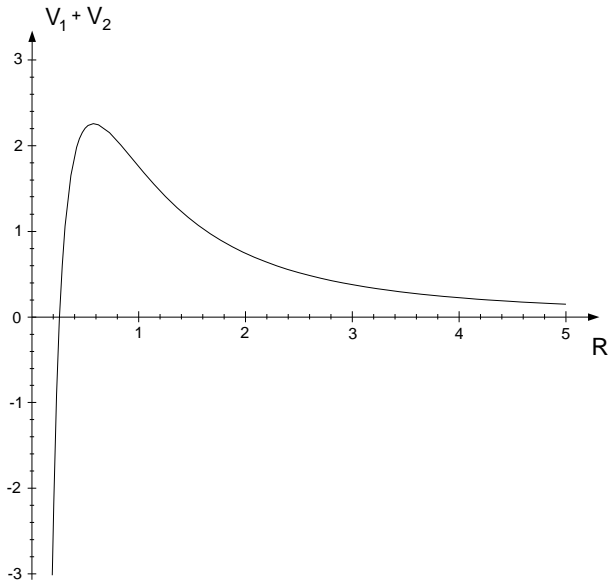}
  \end{center}
\caption{The bound-state potential $V_1 + V_2$ as a function of $R =
r/\rho$ for the attractive case (quantum numbers $M = m = 1/2$).}
\label{fig:POTENTIAL}
\end{figure}
For $M = 1/2$, the normalizable solution of (\ref{GEQ2}) is given by
\be
  g(R) = \sqrt{R} \left[ 1 - (1 + R^2) \ln \left( 1 + \frac{1}{R^2}
  \right) \right] \; .
\ee
Close to the origin, $f(R) = g(R)/R^{3/2}$ behaves as 
\be
  f(R) =  \frac{1}{R} (1 + 2 \ln R) - R (1 - 2 \ln R) + O(R^2) \; , 
\ee
while asymptotically it drops as $1/R^3$. Both types of behavior are
sufficient to make $f$ (or $\phi$) normalizable.  The radial wave
function $f(R)$ and the associated probability distribution $p(R) =
R^3 f^2 (R)$ are shown in Fig.~\ref{fig:FW}.
\begin{figure}[t]
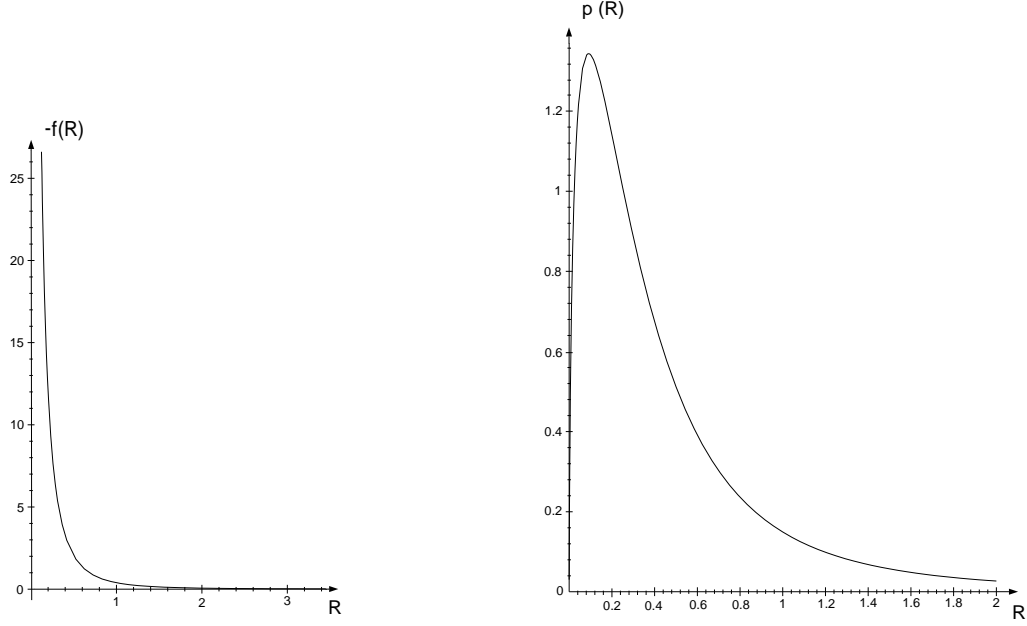

%\unitlength=1pt
\begin{minipage}{0.5\linewidth}
\begin{picture}(190,235)
\put(-20,-43){\includegraphics{radzm.epsf}}
\end{picture}
\hfill
\end{minipage}
\begin{minipage}{0.5\linewidth}
\begin{picture}(190,235)
\put(-25,-40){\includegraphics{radprob.epsf}}
\end{picture}
\end{minipage}
\caption{The radial wave function $- f(R)$ of the zero mode and the
associated probability distribution $p(R)$. While $f$ diverges at the
origin, $p$ vanishes due to the radial measure factor $R^3$.}
\protect{\label{fig:FW}}
\end{figure}
From this figure it is obvious that $f$ has no nodes and therefore
corresponds to the ground state in the sector with $M = 1/2$ (cf.~the
analogous reasoning in \cite{vanbaal:92}).
 
The degeneracy of the solution is found as follows. FP does not depend
on $N_3$, therefore $n$ can arbitrarily be chosen as an half--integer
from $\{ -M, -M + 1, \ldots, M \}$, i.e.~for $M$ = 1/2, one has $n$ =
$\pm$ 1/2. Furthermore, FP is invariant under $(m,s) \to (-m, -s)$, so
that, altogether, there is a four--fold degeneracy. In terms of
abstract states $|M, m, n, s \ket$ the zero modes are linear
combinations of the four degenerate basis states $|1/2 , 1/2 , \pm 1/2
, - \ket $ and $|1/2 , - 1/2 , \pm 1/2 , + \ket $.

\begin{figure}
  \begin{center}
    \epsfig{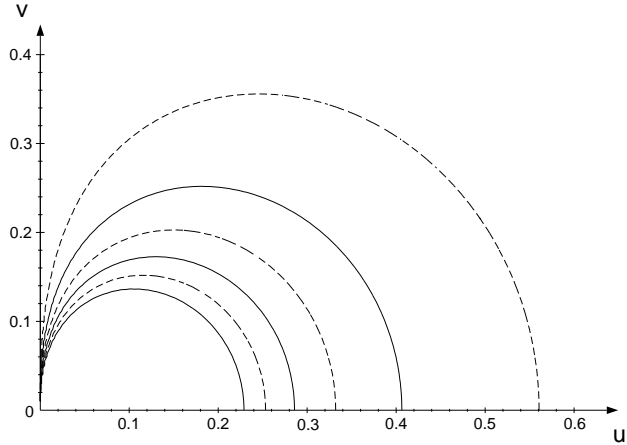}
  \end{center}
\caption{Lines of constant modulus $\varphi$ of the zero mode $\phi
\equiv \Psi_4$ as a function of $u \equiv R \cos \theta$ and $v \equiv R 
\sin \theta$. The dashed lines correspond to $\varphi =
\pi/2$, $3\pi/2$,  $5\pi/2$, ..., the full ones to $\varphi = \pi$, $2\pi$,
$3\pi$, ..., with $\varphi$ increasing from the outermost curve
towards the origin.}
\label{fig:ISOHIGGS}
\end{figure}

To explicitly determine the zero mode, we still have to find the
functions $h_{Mmn}$, $y_{mn}$ and $z_{mn}$ for $M$ = 1/2.  $y_{mn}$
and $z_{mn}$ are eigenfunctions of the operators $L_3$ and $K_3$ so
that their product becomes an eigenfunction of the two operators
\bea
  M_3 &=& \frac{1}{2} (L_3 - K_3) = - \frac{i}{2} \left(
  \pad{}{\varphi_{12}} - \pad{}{\varphi_{34}} \right) \; , \\
  N_3 &=& \frac{1}{2} (L_3 + K_3) = - \frac{i}{2} \left(
  \pad{}{\varphi_{12}} + \pad{}{\varphi_{34}} \right) \; , 
\eea
according to 
\bea
  M_3 \, y_{mn} \, z_{mn} &=& m \, y_{mn} \, z_{mn} \; , \\
  N_3 \, y_{mn} \, z_{mn} &=& n \, y_{mn} \, z_{mn} \; .
\eea
Explicitly, one finds
\bea
  y_{mn} (\varphi_{12}) &=& \e^{i (m+n) \varphi_{12}} \; , \\
  z_{mn} (\varphi_{34}) &=& \e^{- i (m-n) \varphi_{34}} \; . 
\eea   
The function $h_{Mmn} (\theta)$ satisfies the differential equation
\be
\label{THETA_EQ}
  \left[ \frac{1}{\sin 2\theta} \pad{}{\theta} \sin 2\theta
  \pad{}{\theta} + 4M (M+1) - \frac{(m+n)^2}{\cos^2 \theta} -
  \frac{(m-n)^2}{\sin^2 \theta} \right] h_{Mmn} (\theta) = 0 \; .
\ee
For $M$ = 1/2, we can circumvent solving this equation by
considering only the two extremal states in a multiplet with  $m =
\pm M$, which obey
\be
  M_\pm |M, \pm M, n \ket = 0 \; .
\ee
The associated differential equation is much simpler than
(\ref{THETA_EQ}) and straightforwardly solved in terms of the
functions
\bea
  h_{M, -M, n} (\theta) &=& \cos^{M-n} \theta \, \sin^{M+n} \theta \; ,
  \nn \\
  h_{M,  M, n} (\theta) &=& \sin^{M-n} \theta \, \cos^{M+n} \theta \; .
\label{HMMN}
\eea
Direct application to $m = \pm 1/2$ finally yields the four degenerate
zero modes for $M$ = 1/2 (using the notation $\phi_{mns}$),
\be
  \begin{array}{rclcl}
  \phi_{-1/2, -1/2, +} (x) &=& c f(r) \cos \theta \, \e^{-i \varphi_{12}}
  \chi_+ &\equiv& \Phi_1 \; ,   \\ 
  \phi_{-1/2, 1/2, +} (x) &=& c f(r) \sin \theta \, \e^{i \varphi_{34}}
  \chi_+ &\equiv& \Phi_4 \; ,  \\ 
  \phi_{1/2, 1/2, -} (x) &=& c f(r) \cos \theta \, \e^{i \varphi_{12}}
  \chi_- &\equiv& \Phi_2 \; ,  \\ 
  \phi_{1/2, -1/2, -} (x) &=& c f(r) \sin \theta \, \e^{-i \varphi_{34}}
  \chi_- &\equiv& \Phi_3 \; , 
  \end{array}  \label{4MODES}
\ee
where $c$ denotes a normalization constant which will be determined in
a moment. To this end we rewrite the measure 
\be
  d^4 x = r^3 dr \cos \theta \sin \theta \, d\theta \, d\varphi_{12} \,
  d\varphi_{34} \; ,
\ee
and calculate the integral ($\Phi$ denoting any of the basic zero modes)
\be
  \int d^4 x \; \Phi^* (x) \cdot \Phi (x) = c^2 \rho^4 \, \frac{\pi^2}{6}
  \left(1 + \frac{\pi^2}{3} \right) \stackrel{!}{=} 1 \; .
\ee
This determines the normalization $c$. Any zero mode $\phi$ of FP
satisfying (\ref{FPZM}) must be a linear combination of the four basis
modes (\ref{4MODES}). For the following considerations it is convenient
to introduce the real basis,
\renewcommand{\arraystretch}{1.8}
\newcommand{\DS}{\displaystyle}
\be
  \begin{array}{rclcc}
  \Psi_1 &\equiv& \frac{1}{2i} (\Phi_3 - \Phi_4) &=& \frac{c}{\sqrt{2}}
  \frac{f(r)}{r}  \left[\DS -x_4 \atop \DS -x_3 \right] \; ,  \\
  \Psi_2 &\equiv& \frac{1}{2} (\Phi_3 + \Phi_4) &=& \frac{c}{\sqrt{2}}
  \frac{f(r)}{r}  \left[\DS x_3 \atop \DS -x_4 \right] \; ,  \\
  \Psi_3 &\equiv& \frac{1}{2i} (\Phi_1 - \Phi_2) &=& \frac{c}{\sqrt{2}}
  \frac{f(r)}{r}  \left[\DS -x_2 \atop \DS x_1 \right] \; ,  \\
  \Psi_4 &\equiv& \frac{1}{2} (\Phi_1 + \Phi_2) &=& \frac{c}{\sqrt{2}}
  \frac{f(r)}{r}  \left[\DS x_1 \atop \DS x_2 \right] \; , 
  \end{array}
\ee
which, upon using the properties of `t~Hooft's $\eta$ symbols
\cite{schaefer:98} can be compactly written as
\be
  \Psi_\mu^{\bar a} (x) = \frac{c}{\sqrt{2}} \, f(r) \, \bar \eta_{\mu
  \nu}^{\bar a} \, \hat x_\nu \; .
\ee
A general linear combination thus assumes the form
\be
\label{LCZM}
  \phi^{\bar a} (x) \equiv  \frac{\sqrt{2}}{c} n_\mu \Psi_\mu^{\bar a}
  = n^\mu \, \bar \eta_{\mu \nu}^{\bar a} \,
  \hat{x}^\nu \, f(r) \equiv m^{\bar a} f(r)/r \; ,
\ee
where $n_\mu$ is a constant four vector. The latter is particularly
suited for obtaining the finite transformation, 
\be
\label{FINZM}
  \Omega = \exp i \phi^{\bar a} \sigma^{\bar a}/2 = \Eins \cos
  \varphi/2 + i N^{\bar a} \sigma^{\bar a} \sin \varphi/2 \; , 
\ee
with $\varphi = (\phi^{\bar a} \phi^{\bar a})^{1/2}$ and $N^{\bar a} =
\phi^{\bar a}/ \varphi$. Using (\ref{LCZM}) one finds the explicit
representation 
\bea
  \varphi &=& \frac{f(r)}{r} \sqrt{m^{\bar a} m^{\bar a}} \; , \\
  N^{\bar a} &=& \frac{m^{\bar a}}{\sqrt{m^{\bar a} m^{\bar a}}} \; .
\eea
Applying the gauge transformation $\Omega$ from (\ref{FINZM}) to the
singular instanton leads to a configuration that is no longer in the
MAG. This is at variance with the solution $\Omega_R$ found by Brower et
al. \cite{brower:97b} which yields a monopole configuration
\textit{within} the MAG. To illustrate this difference we plot the
modulus $\varphi$ (denoted $\beta$ in \cite{brower:97b}) for the
choice $\phi = \Psi_4$ or, correspondingly, $n = (0,0,0,1)$, $\vc{m} =
(x_1, x_2)^T$. The result is shown in Fig.~\ref{fig:ISOHIGGS} which
clearly differs from the analogous Fig.~2 in \cite{brower:97b}.
The presence of a zero mode as given by (\ref{LCZM}) shows that the
instanton in the singular gauge is located on the Gribov horizon of
the MAG. For (covariant) background type gauges, an analogous result
has been obtained in \cite{baulieu:96}.

\section{Discussion}  

Among the different Abelian gauges used for the lattice study of the
dual superconductor hypothesis, the MAG is the one that has been
analysed in greatest detail. In this paper we have tried to supplement
these achievements by analytic investigations. As the gauge fixing is
nonlinear, this requires some effort. We have calculated the FP
operator for general gauge group $SU(N)$. The result is fairly
complicated; considerable simplifications only seem to arise for gauge
group $SU(2)$. For this particular case we were able to show by quite
general reasoning that there must be Gribov copies. This finding was
confirmed both for a simple toy model and the full field theory. In
the latter case it turns out that the singular instanton is a horizon
configuration in the MAG. The associated zero modes of the FP operator
have explicitly been constructed.

Let us finally discuss some possible physical consequences of our
findings. The two pronounced manifestations of the QCD vacuum are
confinement and spontaneous breakdown of chiral symmetry.  As stated
above, the MAG is well suited for studying the former by checking dual
superconductivity which is believed to be due to monopole condensation
\cite{parisi:75,mandelstam:76,thooft:76a}. On the lattice, condensation
of monopoles has been confirmed for various Abelian gauges
\cite{deldebbio:95,nakamura:97,chernodub:97,digiacomo:99}. The 
monopole vacuum, however, does not provide a straightforward
explanation of chiral symmetry breaking which is due to instantons
rather than monopoles \cite{thooft:76c,diakonov:95,schaefer:98}.  It
is thus of conceptual importance to relate these two complementary
pictures of the vacuum to each other. In computer `experiments'
correlations between instantons and monopoles have indeed been
detected
\cite{goeckeler:89,suganuma:95,bornyakov:96,hart:96,brower:97b,thurner:96,suganuma:97,fukushima:97,sasaki:99}.
The dynamical origin of these correlations, however, remains unclear,
despite considerable efforts to investigate this problem analytically,
in the MAG \cite{chernodub:95,brower:97b,brower:97a,sasaki:99}, the
Polyakov gauge
\cite{reinhardt:97b,jahn:98,ford:98,ford:99a,ford:99b} and
other Abelian gauges \cite{ilgenfritz:00,jahn:99}.  For the MAG, the
situation is as follows. There are basically three known solutions
which represent \emph{finite} transformations from the singular
instanton $A^{\mathrm{sing}}$ into another MAG configuration. These
are (i) the transformation (\ref{SINGREG}) to the regular gauge
instanton $A^{\mathrm{reg}}$, (ii) the `hedgehog' transformation of
Chernodub and Gubarev
\cite{chernodub:95}, and (iii) the family of solutions $\{A(R)\}$
given by Brower et al.~\cite{brower:97b}, interpolating between
$A^{\mathrm{sing}}$ and $A^{\mathrm{reg}}$.  Of these solutions only
(ii) and (iii) induce magnetic monopoles. Solution (ii) leads to an
infinite Dirac string, solution (iii) to a monopole loop of radius
$R$. The associated MAG functional $F$ diverges in cases (i) and
(ii). In case (iii) it is finite, however such that $F[A(R)] > F[A(0)]
\equiv F[A^{\mathrm{sing}}]$. As a result one concludes that the
instanton in the singular gauge defines the global minimum of $F$
along the single instanton orbit. In other words, the MAG functional
does not support monopoles associated with single instantons as these
configurations give rise to a larger value of $F$. This is actually
consistent with lattice results. In \cite{hioki:91a,bali:96,hart:97} it was
observed that the number of monopoles decreases the better the MAG is
fixed, i.e.~the closer one approaches the absolute maximum of the
lattice MAG functional. Due to monopole dominance, the string tension
also becomes smaller. This effect might well be due to the suppression
of monopole loops associated with single instantons.

In favor of the instanton--monopole correlation, Brower et al.~argue
that a possible zero mode of FP can be interpreted as a kinematical
instability of the singular instanton against monopole formation. In
the limit of small monopole loops, $R \ll \rho$, their solution
(eq.~(31) in
\cite{brower:97b}) indeed is a zero mode of FP. It goes like $\sin
2\theta \sim \sin \theta \cos \theta$, and thus, upon comparing with
(\ref{HMMN}), is seen to correspond to $M=1$. Therefore, from our
general analysis in the preceding section, it is \textit{not
normalizable} and thus should be discarded from the stability
analysis. It is probably not too surprising that singular gauge
transformations like the ones found in \cite{brower:97b} lead to zero
modes with diverging norm. 

The physical interpretation of the normalizable $M = 1/2$ zero mode
given in (\ref{LCZM}) is not completely clear. We have checked that it
is not due to any of the known space-time symmetries of
the instanton. Contrary to our expectations, it also has nothing to do
with the solution of Brower et al. In particular, it does \textit{not}
induce monopole singularities. Furthermore, as stated in the last
section, the finite transformation (\ref{FINZM}) even leads out of
the MAG. All this confirms the result that in the MAG \textit{single}
instantons are not correlated with monopoles. One is thus left with a
possible correlation between \textit{multi}-instanton configurations
and monopoles. Numerically, this has been observed
\cite{goeckeler:89,suganuma:95,bornyakov:96,hart:96,brower:97b,thurner:96}.
In particular, the instanton-anti-instanton (IA) system seems to be
physically interesting. In this case one finds that both I and A are
surrounded by a single monopole loop if the IA distance is
large. Below a critical distance, however, the two loops merge into a
single one \cite{brower:97b} which can be viewed as a `kinematical
precursor' to monopole percolation. Of course, an analytic treatment
of multi-instanton systems is quite involved, but maybe not
hopeless. In this respect let us just mention Rossi's old construction
of the BPS monopole in terms of an infinite number of instantons
aligned along the time axis \cite{rossi:79}. We have performed some
preliminary investigations of the IA system which show that the simple
sum ansatz, $A^{IA} = A^I + A^A$ is not in the MAG. The ansatz
suggested by Yung
\cite{yung:88}, however, does fulfill the differential MAG conditions
(\ref{MAGFIX}), though the MAG functional probably diverges. Further
work in this direction is surely necessary.

\section*{Acknowledgements}

The authors thank M.~M\"uller-Preussker for enlightening discussions
and D.~Han\-sen for a careful reading of the manuscript. T.H.~and
T.T.~acknowledge support under DFG grant Wi-777 and RE 856/4-1,
respectively. Part of the research of T.T.~was performed during his
stay at the Institute of Theoretical Physics in Jena.

\appendix

\section{Notations and Conventions}
\label{app:A}

The generators of $SU(N)$ are \emph{hermitean} matrices denoted by
$T^a$ with normalization $\tr(T^a T^b) = \delta^{ab}/2$.  Any gauge
field $A = A^a T^a$ is decomposed into a component $A^\parallel$ in
the Cartan subalgebra $\mathcal{H}^\parallel \subset su(N)$ and a
component $A^\perp$ in the complement, $\mathcal{H}^\perp$, such that
$su(N) = \mathcal{H}^\parallel \oplus \mathcal{H}^\perp$ and
\be
\label{CARTAN_DECOMP}
  A = A^\parallel + A^\perp = A^i H_i + A^\alpha E_\alpha \; .
\ee
The different generators obey the commutation relations \cite{wybourne:74}
\bea
\label{BASIS}
  [H_i , H_j] &=& 0 \; , \quad  i = 1, \ldots, r \; , \\ ~
  [H_i , E_\alpha] &=& \alpha_i E_\alpha \; , \\ ~
  [E_\alpha , E_\beta] &=& N_{\alpha \beta} E_{\alpha + \beta} \; ,
  \quad \alpha + \beta \ne 0 \; , \label{OBS} \\ ~
  [E_\alpha, E_{-\alpha}] &=& \alpha_i H_i \; .
\eea
The rank of the Lie algebra is denoted by $r$, the $\alpha_i$ are the
roots, and $N_{\alpha \beta}$ is a normalization the value of which is
not important for us. For $SU(2)$, which has only two roots $\pm \alpha$,
the third commutator (\ref{OBS}) becomes obsolete, and the situation
simplifies considerably.

The decomposition (\ref{CARTAN_DECOMP}) is orthogonal with respect to
the scalar product
\be
\label{YM_SCAL_PROD} 
  \bra A, B \ket \equiv \int d^d x \, \tr A B \; , 
\ee
where $A$ and $B$ denote some arbitrary Lie algebra valued
$\mathbb{L}_2$ functions. Thus we have 
\be
  \bra A^\perp , B^\parallel \ket = 0 \; . 
\ee
We will also use an alternative notation \cite{quandt:98a} where we
simply divide the $N^2 -1$ generators $T^a$ into 'neutral' and
`charged' ones by means of their superscripts, namely $T^a = (T^{a_0},
T^{\bar a})$ with $T^{a_0} \in \mathcal{H}^\parallel$ and $T^{\bar
a} \in \mathcal{H}^\perp$. A gauge field thus is decomposed as
\be
\label{REIN_NOT}
  A_\mu = A_\mu^a \, T^a = A_\mu^{a_0} \, T^{a_0} + A_\mu^{\bar a} \,
  T^{\bar a} \; . 
\ee
The superscripts $a_0$ and $\bar a$ take on $r$ and $N^2 - 1 - r$
values, respectively. For $SU(2)$, for example, we have $a_0$ = 3 and
$\bar a \in \{1,2\}$, while for $SU(3)$, $a_0 \in \{3,8 \}$ etc.

\section{Group-Theoretical Identities}
\label{app:B}

In this appendix we prove the two useful identities,
\bea
  U^{-1} A_\mu U &=& \exp(i \ad \phi) A_\mu \; , \label{GAUGETR_EXPL1}
  \\ 
  i U^{-1} \partial_\mu U &=& \frac{\exp(i \ad \phi) - \Eins}{i \ad
  \phi} \, \partial_\mu \phi \label{GAUGETR_EXPL2} \; ,
\eea
which hold for an arbitrary gauge transformation $U = \exp (-i \phi)$.
In the above, we have denoted $\ad (A) B \equiv [A, B]$.

(\ref{GAUGETR_EXPL1}) is simply the definition of the adjoint
representation of a Lie group expressed in terms of the adjoint
representation of the Lie algebra \cite{cornwell:84},
\be
  \exp(i \phi) X \exp (- i \phi) \equiv \Ad \Big( \exp (i \phi) \Big)
  X = \exp(i \ad \phi) X \; ,
\label{ADJOINT_REP}
\ee
where $X$ is an arbitrary Lie algebra element. Equation
(\ref{GAUGETR_EXPL2}) is obtained from the identity
\be
  i \exp (is \phi) \partial_\mu \exp(-is \phi) = \frac{\exp(i s \ad
  \phi) - \Eins}{i \ad \phi} \partial_\mu \phi \; ,
\label{LEMMA1}
\ee
for $s=1$. To show (\ref{LEMMA1}) we first note that it is obviously
true for $s = 0$. Differentiating with respect to $s$, we find
\bea
  \pad{}{s} \mbox{l.h.s.} &=& \exp(is\phi) (\partial_\mu \phi)
  \exp(-is\phi) \; , \nn \\
  \pad{}{s} \mbox{r.h.s.} &=& \exp(is \ad \phi) \partial_\mu \phi = 
  \exp(is\phi) (\partial_\mu \phi)
  \exp(-is\phi) \; , 
\eea
where in the last step we have used (\ref{ADJOINT_REP}). Upon
inspection we note that both sides of (\ref{LEMMA1}) obey the same
first order differential equation in $s$ and initial condition at
$s=0$. Thus (\ref{LEMMA1}) is true for all $s$.

\section{The Second Derivative of the MAG Functional}
\label{app:C}

In this appendix we calculate the second derivative of the MAG
functional given by the last two terms in (\ref{FTAYLOR}).  First we
evaluate $(D_\mu \phi)^\perp$,
\bea
  (D_\mu \phi)^\perp &=& \partial_\mu \phi^\perp - i [A_\mu^\parallel ,
  \phi^\perp] - i [A_\mu^\perp , \phi^\perp]^\perp - i [A_\mu^\perp,
  \phi^\parallel] \nn \\
  &=& (D_\mu \phi^\perp)^\perp -i [A_\mu^\perp, \phi^\parallel] \; .
\eea
This yields for  the square term in (\ref{FTAYLOR}),
\bea
\label{TERM21}
  \| (D_\mu \phi)^\perp \|^2 &=& \| (D_\mu \phi^\perp)^\perp - i
  \, [A_\mu^\perp , \phi^\parallel] \|^2 \nn \\
  &=& \bra D_\mu \phi^\perp , (D_\mu \phi^\perp)^\perp \ket - 2i \, \bra
  D_\mu \phi^\perp , [A_\mu^\perp , \phi^\parallel] \ket - \bra
  [A_\mu^\perp , \phi^\parallel], [A_\mu^\perp , \phi^\parallel] \ket
  \nn \\
  &=& - \bra \phi^\perp , D_\mu \mathbb{Q} D_\mu \phi^\perp \ket + 2i \,
  \bra \phi^\perp , [A_\mu^\perp , D_\mu \phi^\parallel] \ket + \bra
  \phi^\parallel , [A_\mu^\perp , [A_\mu^\perp , \phi^\parallel]] \ket
  \nn \\
  &+& 2 i \, \bra \phi^\perp , [D_\mu A_\mu^\perp, \phi^\parallel] \ket \;
  .
\eea
In the last equality, we have made use of the `Leibniz rule',
\be
\label{LEIBNIZ}
  D_\mu [B, C]  =  [D_\mu B , C]  +  [B, D_\mu C]  \; , 
\ee
and defined a projection $\mathbb{Q}$ onto the Cartan complement,
$\mathbb{Q} A = A^\perp$. Note that the last term in (\ref{TERM21})
vanishes at the critical points (\ref{MAGF}). 
The second term of order $\phi^2$ in (\ref{FTAYLOR}) is 
\bea
\label{TERM22}
  i \bra A_\mu^\perp , [\phi , D_\mu \phi]^\perp \ket &=& - i \bra
  \phi , [A_\mu^\perp , D_\mu \phi] \ket \nn \\
  &=& -i \bra \phi^\perp , [A_\mu^\perp , D_\mu \phi^\perp] \ket -i
  \bra \phi^\perp , [A_\mu^\perp , D_\mu \phi^\parallel] \ket \nn \\
  && -i \bra \phi^\parallel , [A_\mu^\perp , D_\mu \phi^\perp] \ket -i
  \bra \phi^\parallel , [A_\mu^\perp , D_\mu \phi^\parallel] \ket \; .
\eea
The third term can be reshuffled and evaluated with the rule
(\ref{LEIBNIZ}) yielding
\be
  -i \bra \phi^\parallel , [A_\mu^\perp , D_\mu \phi^\perp] \ket = -i
  \bra \phi^\perp , [D_\mu A_\mu^\perp , \phi^\parallel] \ket -i \bra
  \phi^\perp , [A_\mu^\perp , D_\mu \phi^\parallel] \ket \; .
\ee
Plugging this into (\ref{TERM22}) and adding (\ref{TERM21}) we see
that the terms which mix $\phi^\perp$ and $D_\mu \phi^\parallel$
cancel. The $O(\phi^2)$ term in $F$ thus becomes
\bea
  F^{(2)} [A; \phi] &\equiv& - \bra \phi^\perp , D_\mu \mathbb{Q} D_\mu
  \phi^\perp \ket - i \, \bra \phi^\perp , [A_\mu^\perp, D_\mu \phi^\perp]
  \ket + i \bra \phi^\perp , [D_\mu A_\mu^\perp, \phi^\parallel] \ket \nn \\
  && + \bra \phi^\parallel , [A_\mu^\perp , [A_\mu^\perp,
  \phi^\parallel]] \ket -i \, \bra \phi^\parallel , [A_\mu^\perp , D_\mu
  \phi^\parallel] \ket  \; .
\label{F20}
\eea
The two terms bilinear in  $\phi^\parallel$ add up to zero according to 
\be
  -i \bra \phi^\parallel , [A_\mu^\perp , (D_\mu + i \ad
  A_\mu^\perp) (\phi^\parallel)] \ket = -i \bra \phi^\parallel ,
  [A_\mu^\perp , D_\mu^\parallel \phi^\parallel] \ket = 
  -i \bra \phi^\parallel , [A_\mu^\perp , \partial_\mu
  \phi^\parallel ] \ket = 0 \; , 
\ee
where the last identity holds because the commutator is in the Cartan
complement $\mathcal{H}^\perp$. Expression (\ref{F20}) thus simplifies
to
\bea
  F^{(2)} [A; \phi] &=& - \bra \phi^\perp , D_\mu \mathbb{Q} D_\mu
  \phi^\perp - i [A_\mu^\perp, D_\mu \phi^\perp] \ket + i \bra
  \phi^\perp , [D_\mu A_\mu^\perp, \phi^\parallel] \ket \nn \\
  &\equiv& F^{(2)} [A; \phi^\perp] + i \bra \phi^\perp , [D_\mu
  A_\mu^\perp, \phi^\parallel] \ket \; .
\label{F21}
\eea
Introducing $\mathbb{P} = \Eins - \mathbb{Q}$, the terms quadratic in
$\phi^\perp$ assume the following form,
\bea 
\label{F22}
  F^{(2)} [A; \phi^\perp] &\equiv& - \bra \phi^\perp , (D_\mu
  \mathbb{Q} + i \ad A_\mu^\perp) (D_\mu \phi^\perp) \ket \nn \\ 
  &=& - \bra \phi^\perp , (D_\mu \mathbb{Q} + i \ad A_\mu^\perp)
  (\mathbb{P} + \mathbb{Q})(D_\mu\phi^\perp) \ket \nn \\
  &=& - \bra \phi^\perp , D_\mu^\parallel \mathbb{Q} \, D_\mu
  \phi^\perp + i \ad A_\mu^\perp \, \mathbb{P} D_\mu \phi^\perp \ket
  \; .
\eea
We thus need the projections 
\bea
  \mathbb{P} D_\mu \phi^\perp &=& -i \mathbb{P} [A_\mu^\perp ,
  \phi^\perp] \; , \\
  \mathbb{Q} D_\mu \phi^\perp &=& D_\mu^\parallel \phi^\perp -i \mathbb{Q}
  [A_\mu^\perp , \phi^\perp] \; .
\eea
Using this and the identity $\bra \phi^\perp , \mathbb{Q}
D_\mu^\parallel A \ket = \bra \phi^\perp , D_\mu^\parallel A \ket$,
(\ref{F22}) becomes
\bea
  && F^{(2)} [A; \phi^\perp] = \nn \\
  &&=  - \bra \phi^\perp , D_\mu^\parallel D_\mu^\parallel
  \phi^\perp -i D_\mu^\parallel \mathbb{Q} [A_\mu^\perp , \phi^\perp]
  + [A_\mu^\perp, \mathbb{P} [A_\mu^\perp , \phi^\perp]] \ket \nn \\
  && = - \bra \phi^\perp , D_\mu^\parallel D_\mu^\parallel \phi^\perp -i
  [ D_\mu^\parallel A_\mu^\perp , \phi^\perp] -i [A_\mu^\perp,
  D_\mu^\parallel \phi^\perp +i \mathbb{P} [A_\mu^\perp , \phi^\perp]]
  \ket \nn \\ 
  && = - \bra \phi^\perp , D_\mu^\parallel D_\mu^\parallel
  \phi^\perp -i [ D_\mu^\parallel A_\mu^\perp , \phi^\perp] -i
  [A_\mu^\perp, D_\mu^\parallel \phi^\perp -i \mathbb{Q} [A_\mu^\perp
  , \phi^\perp] + i [A_\mu^\perp , \phi^\perp]] \ket \nn \\
  && = - \bra \phi^\perp , D_\mu^\parallel D_\mu^\parallel
  \phi^\perp -i [ D_\mu^\parallel A_\mu^\perp , \phi^\perp] -i
  [A_\mu^\perp, \mathbb{Q} D_\mu \phi^\perp]  +  [A_\mu^\perp , [
  A_\mu^\perp, \phi^\perp]] \ket \; . \nn \\
\eea

This is the result used in (\ref{F12}).

\section{The Laplacian of the Toy Model}
\label{app:D}

Using matrix notation, the Laplacian $\Delta^{ab} = D_i^{ac} D_i^{cb}$
of the toy model is given by
\be
  \Delta =  
  \left( \begin{array}{ccc}
           -y^2 - z^2 - Y^2 - Z^2 & xy + XY & xz + XZ \\
           yx + YX & -x^2 - z^2 - X^2 - Z^2 & yz + YZ \\
           zx + ZX & zy + ZY & -x^2 - y^2 - X^2 - Y^2 
         \end{array} \right)  \; .
\ee
Denoting $r \equiv |\vcg{x}|$ and $R \equiv |\vcg{X}|$, the
determinant of the Laplacian becomes
\be 
  \det \Delta = - (r^2 + R^2) \, (\vcg{x} \times \vcg{X})^2 \equiv - (r^2 +
  R^2) \, d^2  \le 0 \; .
\ee
As is appropriate for a Laplacian, $\Delta$ is a negative-semidefinite
operator. It has zero modes for reducible configurations only
\cite{babelon:81b}, which for the case at hand are given by the zero
configuration, $\vcg{x} = \vcg{X} = 0$ (with the full group $SO(3)$ as
its stabilizer), and the collinear configurations, $\vcg{x} = \alpha
\vcg{X}$, with $U(1)$ stabilizer.
                
The eigenvalues of $-\Delta$ are given by 
\bea
  E_0 &=& r^2 + R^2 \; , \\
  E_\pm &=& \frac{1}{2} (r^2 + R^2) \pm \frac{1}{2} \sqrt{(r^2 -
  R^2)^2 + 4 (\vcg{x} \cdot \vcg{X})^2} \; . 
\eea
It is reassuring to note that the eigenvalues and, accordingly, the
determinant only depend on the gauge invariant scalar products $r^2$,
$R^2$ and $\vcg{x} \cdot \vcg{X}$. At the origin, which has the
largest stabilizer, all eigenvalues vanish. For collinear
configurations with $\vcg{x} \cdot \vcg{X} = \pm rR$, the eigenvalues
are $E_- = 0$ and $E_0 = E_+ = r^2 + R^2$, so that there is a zero
mode and the first `excited' state is degenerate. For the horizon
configurations of the MAG, having $\vcg{x} \cdot \vcg{X} = 0$, $r = R
\equiv \rho$, one finds $E_\pm = \rho^2$ and $E_0 = 2 \rho^2$. Thus,
the groundstate becomes degenerate. The latter fact corresponds to the
gauge fixing degeneracies of the \emph{Laplacian gauge}
\cite{schierholz:85,vink:92,sijs:97} as particularly discussed in
\cite{vink:92}. As the MAG and the Laplacian gauge coincide for constant
gauge fields, the degeneracies have to be the same, and, indeed, they are.

\bibliographystyle{h-physrev}
\bibliography{gauge}

\end{document}